\documentclass[12pt,times, english,  a4paper]{article}
\usepackage{times}
\usepackage{graphicx}
\usepackage[latin1]{inputenc}
\usepackage{geometry}
\usepackage{amsthm}
\usepackage{setspace}

\newtheorem{proposition}{Proposition}[section]
\newtheorem{theorem}{Theorem}[section]

\newenvironment{biography}[3][]{%
\footnotesize\unitlength 1mm\bigskip\bigskip\bigskip\parskip=0pt\par%
\rule{0pt}{39mm}\vspace{-39mm}\par%   garantees correct page breaking
\noindent\setbox0\hbox{\framebox(25,32){
#2
}}%   box containing the frame
\ht0=37mm\count10=\ht0\divide\count10 by\baselineskip%  calculates lines
\global\hangindent29mm\global\hangafter-\count10%
\hskip-28.5mm\setbox0\hbox to 28.5mm {\raise-30.5mm\box0\hss}%
\dp0=0mm\ht0=0mm\box0\noindent\bf [#3]\rm}{
\par\rm\normalsize}

\begin{document}

\title{\textbf{Statistical Modelling of Information Sharing: Community, Membership and Content}}

\author{W.-Y. Ng,\hspace{.3cm}W.K. Lin,\hspace{.3cm}D.M. Chiu\\
Department of Information Engineering\\
The Chinese University of Hong Kong\\
\{wyng,wklin3,dmchiu\}@ie.cuhk.edu.hk}
\date{June 29, 2005}

%\begin{singlespace}
\maketitle
%\end{singlespace}
%%
%%%%%%
\begin{abstract}
File-sharing systems, like many online and traditional information sharing
communities (e.g. newsgroups, BBS, forums, interest clubs), are dynamical
systems in nature.  As peers get in and out of the system, the information
content made available by the prevailing membership varies continually in
amount as well as composition, which in turn affects all peers' join/leave
decisions.  As a result, the dynamics of membership and information content
are strongly coupled, suggesting interesting issues about growth, sustenance
and stability.

In this paper, we propose to study such communities with a simple statistical
model of an \textit{information sharing club}.  Carrying their private payloads of
information goods as potential supply to the club, peers join or leave on the
basis of whether the information they demand is currently available.
Information goods are chunked and \textit{typed}, as in a file sharing system where
peers contribute different files, or a forum where messages are grouped by
topics or threads.  Peers' demand and supply are then characterized by
statistical distributions over the type domain.

This model reveals interesting critical behaviour with multiple equilibria. A
sharp growth threshold is derived: the club may grow towards a sustainable
equilibrium only if the value of an \textit{control parameter} is above the
threshold, or shrink to emptiness otherwise.  The control parameter is
composite and comprises the peer population size, the level of their
contributed supply, the club's efficiency in information search, the spread
of supply and demand over the type domain, as well as the goodness of match
between them.

\end{abstract}
%%%%%
\section{Introduction}
The notion of a peer-to-peer system means different things
to different people. To some, it is a way to use the commodity
personal computers to do the job of large and expensive servers
\cite{SETI}, \cite{ll2002}, \cite{hhbb2003}.  Others build application layer multicast
systems out of it \cite{kazza}, \cite{bittorrent}.  But there is one
thing in common for almost all peer-to-peer systems, that is the
coming-together of such a system depends on the number of peers (large
or small) wanting to participate.\\

The formation of such cooperation without central management
has its inherent advantages: it saves the cost of central
management; and more importantly, it automatically adapts to
the need (for example, in terms of time and scope) of the peers
who collectively form the club.
But what are the forces that attract peers
together?  What would cause a peer-to-peer system to grow,
sustain itself, or fall apart?  Are there some fundamental
reasons that apply to different peer-to-peer systems?\\

The purpose of the peer-to-peer systems is invariably to
share some resources or information.  Economists differentiate between two
kinds of goods that are shared: rivalrous and non-rivalrous
goods. The former diminishes when shared.  Compute power,
storage and communication bandwidth are examples of
rivalrous goods.  Many information goods, however,
are inherently non-rivalrous.  In other words, they can be
readily replicated many times with little or no cost.\\

Motivated by the above questions, we formulate a model for
a club where members share non-rivalrous information
goods. Conventional resources such as computers and bandwidth
are assumed to be abundant.  In such a setting, the strength
of a club is determined by the amount as well as the
\emph{composition} of the information content made available by
the club's prevailing membership, and how that content fits
the potential members demands. Based on this simple
model, it is then possible to derive some very basic
conditions for a club to form and sustain.  The model predicts
a critical population size, from which enough peers will
find matching interest and form a club.  Furthermore,
the model can be used to understand the dynamics of
the content and membership, and whether and
how it leads to an equilibrium.\\

Since the model is simple, it is also general enough to be
applied to many other information sharing paradigms.
Examples include web-based collaborative environments,
newsgroups where peers contribute their opinions about different
topics, or other forums or communities for information
sharing.\\

The rest of the paper is organized as follows.  Section 2
is devoted to modeling the peers in terms of their
demand and contribution, which then leads to a model of
a club in terms of its content.
Section 3 models peers' decisions of joining or leaving
a club, and consequently the conditions for club
formation, and other equilibrium properties of it.
Section 4 illustrates the properties of the model through
numerical examples. Section 5 discusses the contribution
of this model, the interpretation of various results,
as well as the limitations of our model.  Finally, we
conclude and discuss future directions.
%%%%%%
\subsection{Related Works}
Many other papers tried to model incentives in
peer-to-peer systems and the resulting club dynamics.
\cite{gj2003} discusses private versus public goods,
and argues that messages shared in web forums are
private goods, thus suggesting sharing is
not simply an altruistic behavior.
Several papers focus on how to relieve the
cost/congestion of some rivalrous resources, such as
bandwidth and other resources that a peer has to consume.
For example \cite{kstt2002} suggests a possible rationale
for peers' contributions is to relieve the bandwidth stress
when they share, their actions thereby benefit the peers
themselves.
\cite{bas2003}, \cite{gbml2001}, \cite{rrsf2003} use
game theoretic approaches to model and understand
the sharing incentives in peer-to-peer networks.
These works also discuss incentive-compatible solutions to
peer-to-peer systems.
In comparison, our model brings out a new angle that is
complementary and somewhat orthogonal to the above works.\\

Our work is in part motivated by \cite{fpcs2004} in which
a general model is used to explain the vitality of a peer-to-peer network
 when different types of
peers are involved.  The type of a peer is characterized by
the peer's generosity, which is used as a threshold to
determine when a peer would contribute to the club
rather than free-ride.  Their model does not explicitly
capture different types of information goods themselves, therefore
the motivation for sharing remains rather abstract.\\

Our work attempts to explain the motivation of the peers by characterizing
the different types of peers based on their contribution and demand of
different types of information goods.  A peer's decision to join a club can
then be related to the extent the club can satisfy the peer's interest
(demand). This sheds more (at least different) insights to what brings peers
together in the first place.
%%%%%%
\section{The Information Sharing Club (ISC) Model} \label{club_model}

The Information Sharing Club (ISC) model has three basic components.
First, a population of $N$ \emph{peers}, denoted by $\mathcal{N}$,
may freely join or leave the club any time at their own will.
Each peer carries a payload of information goods which
are shared with other current members only when he joins the club.\\

Second, information goods are chunked and \emph{typed}, the same way that versions of different
files are served in a file sharing system, or messages of various
topics are hosted in a forum.  Information chunks of the same type are not
differentiated: an instance of information demand specifies
the chunk type only and is satisfied by \emph{any} chunk of that type, as when
request for a file is satisfied with any copy of it, or when information query
returns any piece of information of the specified class (e.g. as
implied by the query criteria, for instance).\\

Third, the club maintains a platform on which information chunks shared by
members are maintained and searched.  A perfect membership system
makes sure that only requests by current members are processed.  A request
may comprise one or more instances of demand, and is successfully
served when all instances are satisfied.  However, the search may not
be perfect and is conducted with efficiency $\rho \in (0, 1]$, defined as the
probability that any shared chunk is actually found in time by the platform in response to a request.\\

We make probabilistic assumptions about both demand and supply:
peer $i$'s demand instances as well as the content of his private payload,
in terms of chunk types, are drawn from statistical distributions.
Specifically, we assume
peer $i$'s private payload comprises $K_i \ge 0$
chunks drawn from distribution $g_i(s),$
$s\in\mathcal{S}\stackrel{\triangle}{=}\{1,2,\ldots\}$ where
$\mathcal{S}$ is the set of all types.
The total payload of any group of members (\emph{membership})
$\mathcal{G}\subset\mathcal{N}$ is then given by
$$
g_{\mathcal{G}}(s) \;\stackrel{\triangle}{=}\;
\frac{\sum_{i\in \mathcal{G}} K_i\;g_i(s)}{\sum_{i\in \mathcal{G}} K_i},\;\;
\mathcal{G}\subset\mathcal{N}
$$
Without loss of generality, we assume the \emph{aggregate supply function}
$g(s)\stackrel{\triangle}{=} g_{\mathcal{N}}(s)$
to be monotonically non-increasing.  The type variable $s$ may then
be interpreted as a \emph{supply rank (s-rank)}.  In other words,
$s=1$ and $s=|\mathcal{S}|$ denote the most and least supplied chunk types
respectively.
Likewise, we define the \emph{aggregate demand function}
$h(s)\stackrel{\triangle}{=}h_{\mathcal{N}}(s)$ where
$$
h_{\mathcal{G}}(s)\stackrel{\triangle}{=}\;
\frac{\sum_{i\in \mathcal{G}} M_i\;h_i(s)}{\sum_{i\in \mathcal{G}} M_i},\;\;
\mathcal{G}\subset\mathcal{N}
$$
as peer $i$ generates demand instances at a rate of $M_i$ chunks per unit
time, drawn from distribution $h_i(s)$, $s\in\mathcal{S}$\footnote{Another
possible ranking of the types is \emph{popularity rank (p-rank)}, which ranks
the types according to the aggregate demand instead. In cases when the p-rank
is more natural to work with, such as when supply is being driven by demand
and p-ranks are more readily known, we may derive the requisite demand
functions in s-rank as
$$
h_i(s) \stackrel{\triangle}{=} \sum_r\frac{\phi(r, s)}{f(r)}\;f_i(r)
$$
where $f_i(r)$ is peer $i$'s demand distribution over the p-rank domain and
$\phi(r,s)$ is the joint distribution of the two rank measures that captures
how well supply follows demand.  (Perfect following would imply
$\phi(r,s)=0\;\forall r\neq s$.)}.\\

For current club membership $\mathcal{C}$, the expected number of chunks of
type $s$ being shared would be given by $\mu_\mathcal{C}(s) \stackrel{\triangle}{=}
n\; k_\mathcal{C} \; g_{\mathcal{C}}(s)$
where $n\stackrel{\triangle}{=}|\mathcal{C}|$ is the membership size and
$k_\mathcal{C}\stackrel{\triangle}{=}
\sum_{i\in\mathcal{C}} K_i/|\mathcal{C}|$ is the payload size
averaged over the current club membership.  Conditioning on the membership
size, we have
$$
\mu_n(s) = n\; k \; g(s)
$$
where $k \stackrel{\triangle}{=} \sum_{i=1}^{N} K_i/N > 0$ is the payload size averaged over all peers.
We assume further that members' contents are drawn independently, which
implies a Poisson distribution for the actual total number of type $s$ chunks
being shared. Subsequently demand instances for chunk type $s$ have an
average failure rate of $e^{-\mu_n(s)\;\rho} = e^{-n \;k \;g(s)\;\rho}.$ The
average success rate of peer $i$'s demand being satisfied in a club of size
$n$ is therefore
\begin{eqnarray}
p_i(n)&\stackrel{\triangle}{=}  E_{h_i(s)}[1-e^{-n \;k \;g(s)\;\rho}]\label{eqn:download_probability}
\end{eqnarray}
where $E[\cdot]$ is the expectation operator. This is compatible with the non-rivalrous assumption as
 it is independent of the level of demand for this chunk type. \\

\subsection{An example: music information sharing club}
Tables \ref{table:supply_example} and \ref{table:demand_example} depict an
example of six peers sharing music information of five different types. For
simplicity, we assume identical payload sizes (identical $K_i$'s) and demand
rates (identical $M_i$'s) so that the aggregate distributions are simple
unweighted averages of the peers' distributions. Table
\ref{table:example_rank} gives the resulting s-ranks and p-ranks of the five
music types. The information may be news and messages about the different
music types when the club is a discussion forum in nature, or musical audio
files when it is a file sharing platform.
\begin{singlespace}
\begin{table}[hbtp]
\begin{center}
\caption{Distributions of peers' private payloads, $g_i(s)$}\label{table:supply_example}
\begin{tabular}{|r|p{1.8cm}|p{1.8cm}|p{1.8cm}|p{1.8cm}|p{1.8cm}|}
\hline
&
Pop&
Classical&
Oldies&
World&
Alternative
\tabularnewline
\hline
\hline
Alfred&
$0.4$&
$0.3$&
$0.1$&
$0.1$&
$0.1$
\tabularnewline
\hline
Bob&
$0.4$&
$0.2$&
$0.2$&
$0.15$&
$0.05$
\tabularnewline
\hline
Connie&
$0.3$&
$0.3$&
$0.2$&
$0.1$&
$0.1$
\tabularnewline
\hline
David&
$0.2$&
$0.3$&
$0.3$&
$0.15$&
$0.05$
\tabularnewline
\hline
Eric&
$0.5$&
$0.05$&
$0.2$&
$0.15$&
$0.1$
\tabularnewline
\hline
Florence&
$0.1$&
$0.4$&
$0.1$&
$0.1$&
$0.3$
\tabularnewline
\hline
\hline
aggregate supply, g(s)&
$0.317$&
$0.258$&
$0.18$&
$0.125$&
$0.12$
\tabularnewline
\hline
\end{tabular}
\end{center}
\end{table}

\begin{table}[hbtp]
\begin{center}
\caption{Distributions of peers demand, $h_i(s)$}\label{table:demand_example}
\begin{tabular}{|r|p{1.8cm}|p{1.8cm}|p{1.8cm}|p{1.8cm}|p{1.8cm}|}
\hline
&
Pop&
Classical&
Oldies&
World&
Alternative
\tabularnewline
\hline
\hline
Alfred&
$0.1$&
$0.4$&
$0.3$&
$0.1$&
$0.1$
\tabularnewline
\hline
Bob&
$0.05$&
$0.5$&
$0.1$&
$0.3$&
$0.05$
\tabularnewline
\hline
Connie&
$0.1$&
$0.2$&
$0.3$&
$0.2$&
$0.2$
\tabularnewline
\hline
David&
$0.1$&
$0.4$&
$0.3$&
$0.15$&
$0.05$
\tabularnewline
\hline
Eric&
$0.1$&
$0.4$&
$0.2$&
$0.2$&
$0.1$
\tabularnewline
\hline
Florence&
$0.2$&
$0.3$&
$0.1$&
$0.2$&
$0.2$
\tabularnewline
\hline
\hline
aggregate demand, h(s)&
$0.108$&
$0.367$&
$0.217$&
$0.192$&
$0.117$
\tabularnewline
\hline
\end{tabular}
\end{center}
\end{table}
\end{singlespace}
\begin{singlespace}
\begin{table}[hbtp]
\begin{center}
\caption{The supply and the popularity rank}\label{table:example_rank}
\begin{tabular}{|r|p{1.8cm}|p{1.8cm}|p{1.8cm}|p{1.8cm}|p{1.8cm}|}
\hline
&
1&
2&
3&
4&
5
\tabularnewline
\hline
\hline
Supply rank ($s$)&
Pop&
Classical&
Oldies&
World&
Alternative
\tabularnewline
\hline
Popularity rank ($r$)&
Classical&
Oldies&
World&
Alternative&
Pop
\tabularnewline
\hline
\end{tabular}
\end{center}
\end{table}
\end{singlespace}
A peer's success rate would depend on the types of goods he demands on one
hand, viz. $h_i(s)$, and the aggregate supply $g(s)$ on the other. For
instance, Alfred's average success rate is given by:
\begin{eqnarray*}
p_{\textrm{\scriptsize{Alfred}}} = 1-(0.1\,(e^{-6\,(0.317)}) + 0.4\,(e^{-6\,(0.258)}) + \ldots + 0.1\,(e^{-6\,(0.12)})) = 0.69\\
\end{eqnarray*}

%%%%%%

\section{Dynamic equilibrium of membership and content}

Generally speaking, peers would join the club as members and share their
private payloads as long as their requests are sufficiently met.  We make two
simplifying assumptions here: (1) a peer would join as long as a single
current request is met, and leave otherwise; and (2) any request comprises
\mbox{$d \ge 1$} instances of demand.  The probability that peer $i$ would
join when membership is $\mathcal{C}$ is then $P_{\mathcal{C},i}
\stackrel{\triangle}{=} {p^d_{\mathcal{C},i}}$ where $p_{\mathcal{C},i}$ is
the probability that an instance of peer $i$'s demand is satisfied when
membership is $\mathcal{C}$. Conditioning on the membership size $n$, the expected joining probability of
peer $i$ is
\begin{equation}
P_i(n) \stackrel{\triangle}{=} P_i(n)^d
\end{equation}

Membership dynamics and content dynamics are closely coupled: as peers join
and leave, they alter the total shared content, inducing others to revise
their join/leave decisions.  The membership size changes always unless the
two-way flows between members and non-members are balanced.

Consequently, we may define a \emph{statistical equilibrium membership size}
$n_{eq}$ as the solution of the balance condition
\begin{eqnarray}
(N-n_{eq}) \bar P(n_{eq})& = &n_{eq} (1 - \bar P(n_{eq})) \nonumber\\
\Leftrightarrow \hspace{1.0in} \bar P(n_{eq}) &= &\frac{n_{eq}}{N} \label{eqn:dynamic_equilibrium}
\end{eqnarray}
where $\bar P(n) = \frac{1}{N} \sum_{i=1}^{N}{P_i(n)}$ is the joining
probability averaged over all peers and all possible memberships of size $n$.
Note that equation (\ref{eqn:dynamic_equilibrium}) is in the form of a fixed
point equation which is indicative of the coupled dynamics of membership and
content.
Further, the stability condition for a fixed point $n_{eq}$ is simply
\begin{equation}
\left.\frac{\partial \bar P(n)}{\partial n}\right|_{n=n_{eq}} < \frac{1}{N} \label{eqn:start_up_condition}
\end{equation}
Note that an empty membership is always a fixed point, and would always be
stable for sufficiently small $N$, in which case autonomous growth from an empty or small membership is very difficult
if not impossible.
\mbox{}\\
\mbox{}\\

\begin{theorem}[Empty Membership Instability]\label{theorem:instability}
Empty membership is unstable if and only if requests are simple,
viz. $d=1$, and% the \emph{composite control parameter} $\pi$\\
\begin{equation}
\pi\;\stackrel{\triangle}{=}\;N \; k \; \rho \sum_s h(s) \; g(s) \ge 1\;\;\;. \label{eqn:critical_condition}
\end{equation}
\end{theorem}
\begin{proof}

Consider:
\begin{eqnarray*}
\hspace{10mm}\bar P(n) \,= \,\frac{1}{N} \sum_{i=1}^{N}{P_i(n)} \,= \,\frac{1}{N}\sum_{i=1}^{N}p_i(n)^d \hspace{10mm}d\ge1\;\;.\\
 \end{eqnarray*}
Differentiating with respect to $n$:
\begin{eqnarray*}
\frac{N}{d}\frac{\partial \bar P(n)}{\partial n} &=&
\sum_{i=1}^N p_i(n)^{d-1}\frac{\partial p_i(n)}{\partial n}\\
\Leftrightarrow\hspace{.5in}
\frac{N}{d k \rho}\frac{\partial \bar P(n)}{\partial n} &=&
\sum_{i=1}^N p_i(n)^{d-1} E_{h_i(s)}[e^{-n k \rho g(s)}g(s)]
\end{eqnarray*}
Since $p_i(0) = 0$,  it follows that $\left.\partial \bar P(n)/\partial n\,\right|_{n=0} = 0$ for $d > 1$,
in which case an empty membership is always stable.  When $d=1$,
\begin{eqnarray*}
\frac{N}{k \rho}\frac{\partial \bar P(n)}{\partial n} &=&\sum_{i=1}^N E_{h(s)}\,[g(s)] =N \sum_s h(s) g(s)\\
\Leftrightarrow\hspace{.5in} \frac{\partial \bar P(n)}{\partial n} &=& k \rho \sum_s \;h(s)g(s)
\end{eqnarray*}
whence (\ref{eqn:critical_condition}) follows from the stability condition (\ref{eqn:start_up_condition}) for the empty membership fixed point $n_{eq} = 0$.
\end{proof}

\vspace{0.3cm}
%%%%%%%
%%20050301 wyng

In our model, we regard empty membership instability as a necessary condition for autonomous growth from an
empty or small club membership.  The above theorem implies that favourable
conditions are large $k$ (contribution from members),
large $\rho$ (search efficiency) and
a large value of $\sum_s h(s)g(s)$, an inner product of $h(s)$ and $g(s)$.
Note that
$$
\sum_s h(s)g(s)\equiv \Vert h \Vert\,\Vert g \Vert \cdot \langle h(s),g(s)\rangle
$$
where $\Vert h \Vert$ and $\Vert g \Vert$ are the 2-norms of
$h(s)$ and $g(s)$ respectively, and \mbox{$\langle h(s),g(s)\rangle$} is their normalized
inner product which measures their similarity, or goodness of match.
Other favourable conditions are therefore a good match between
aggregate demand and supply, and \emph{skewness} -- or
small spread -- of their distributions over the chunk types.
%%%%%%
%20050226
\subsection{Music information sharing club example with simple requests}

\begin{figure}[h!]
\begin{center}
{\scriptsize
\begin{picture}(0,0)%
\includegraphics{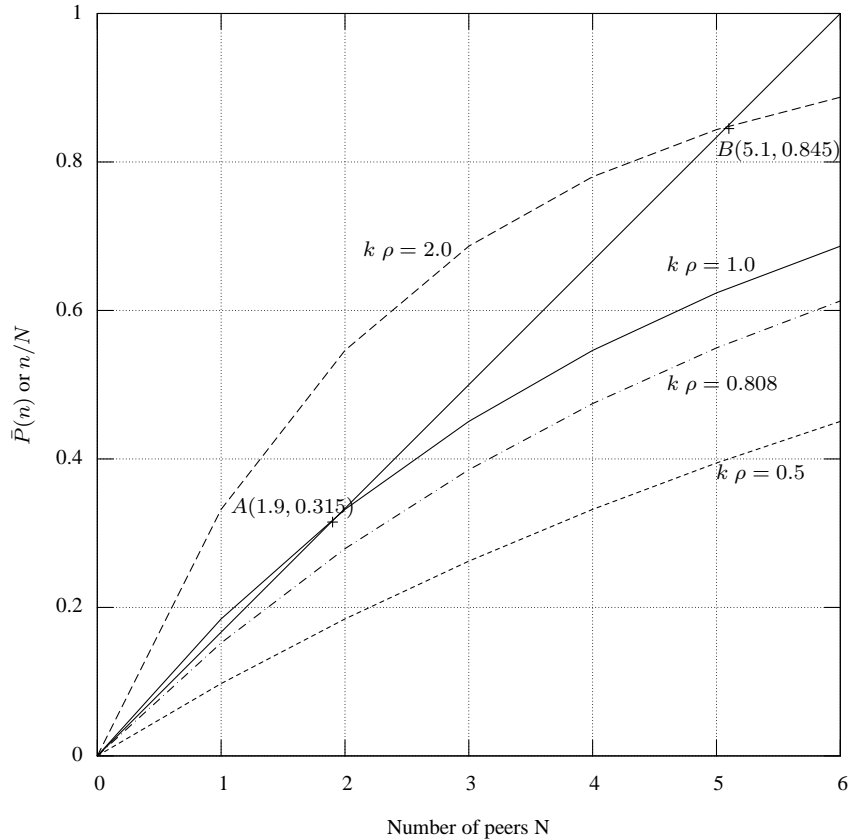}%
\end{picture}%
\begingroup
\setlength{\unitlength}{0.0200bp}%
\begin{picture}(16200,16200)(0,0)%
\put(1650,1650){\makebox(0,0)[r]{\strut{} 0}}%
\put(1650,4450){\makebox(0,0)[r]{\strut{} 0.2}}%
\put(1650,7250){\makebox(0,0)[r]{\strut{} 0.4}}%
\put(1650,10050){\makebox(0,0)[r]{\strut{} 0.6}}%
\put(1650,12850){\makebox(0,0)[r]{\strut{} 0.8}}%
\put(1650,15650){\makebox(0,0)[r]{\strut{} 1}}%
\put(1925,1100){\makebox(0,0){\strut{} 0}}%
\put(4258,1100){\makebox(0,0){\strut{} 1}}%
\put(6592,1100){\makebox(0,0){\strut{} 2}}%
\put(8925,1100){\makebox(0,0){\strut{} 3}}%
\put(11258,1100){\makebox(0,0){\strut{} 4}}%
\put(13592,1100){\makebox(0,0){\strut{} 5}}%
\put(15925,1100){\makebox(0,0){\strut{} 6}}%
\put(550,8650){\rotatebox{90}{\makebox(0,0){\strut{}$\bar P(n)$ or $n/N$}}}%
\put(8925,275){\makebox(0,0){\strut{}Number of peers N}}%
\put(4445,6340){\makebox(0,0)[l]{\strut{}$A(1.9, 0.315)$}}%
\put(13592,13060){\makebox(0,0)[l]{\strut{}$B(5.1, 0.845)$}}%
\put(6942,11170){\makebox(0,0)[l]{\strut{}$k\;\rho = 2.0$}}%
\put(12658,10890){\makebox(0,0)[l]{\strut{}$k\;\rho = 1.0$}}%
\put(12658,8650){\makebox(0,0)[l]{\strut{}$k\;\rho = 0.808$}}%
\put(13592,6970){\makebox(0,0)[l]{\strut{}$k\;\rho = 0.5$}}%
\end{picture}%
\endgroup
} \caption{The music information sharing club example}
\label{figure:club_dynamics}
\end{center}
\end{figure}

Figure (\ref{figure:club_dynamics}) shows $\bar P(n)$ for the music
information sharing club example for four $k\rho$ values for the simple
request case, viz. $d=1$.

For $k\rho=2$, the model predicts that an empty club is unstable.
Any disturbance, e.g. voluntary sharing or contribution, would trigger it
to grow.  The club would stagger rapidly towards the fixed point
$n=5.1$ ---
where $\bar P(x)=5.1/6=0.85$ and and sustain itself around there.
The peers are active members over $80\%$ of the time on average.
For $k\rho=1$, an empty club is again unstable but the club sustains
itself at a smaller average size of $n=1.9$.  With less supply and/or less
efficient search function, peers are active only around $30\%$ of the time on average.
For $k\rho=0.5$, an empty club now becomes stable.  Joining peers are
always more than offset by leaving members such that a positive membership
is always  transient.  Peers are almost always inactive.
Finally $k\rho=(N \sum_s h(s) g(s))^{-1}=0.808$ is the critical case when
an empty club is just stable/unstable.\\

\begin{figure}[h!]
\begin{center}
{\scriptsize
    \begin{picture}(0,0)%
\includegraphics{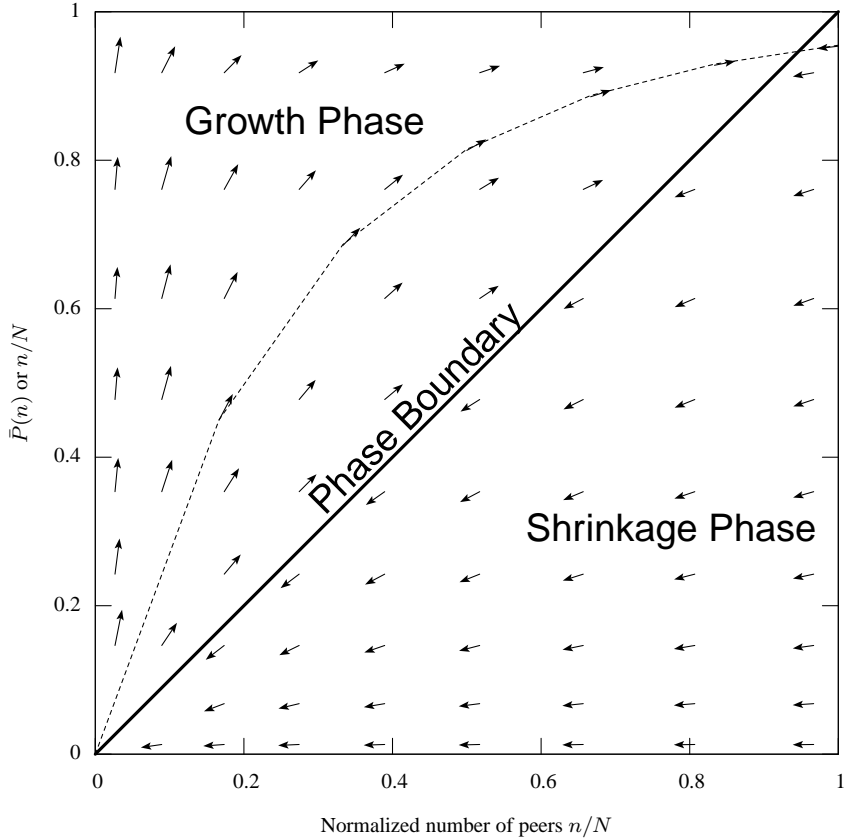}%
\end{picture}%
\begingroup
\setlength{\unitlength}{0.0200bp}%
\begin{picture}(16200,16200)(0,0)%
\put(1650,1650){\makebox(0,0)[r]{\strut{} 0}}%
\put(1650,4450){\makebox(0,0)[r]{\strut{} 0.2}}%
\put(1650,7250){\makebox(0,0)[r]{\strut{} 0.4}}%
\put(1650,10050){\makebox(0,0)[r]{\strut{} 0.6}}%
\put(1650,12850){\makebox(0,0)[r]{\strut{} 0.8}}%
\put(1650,15650){\makebox(0,0)[r]{\strut{} 1}}%
\put(1925,1100){\makebox(0,0){\strut{} 0}}%
\put(4725,1100){\makebox(0,0){\strut{} 0.2}}%
\put(7525,1100){\makebox(0,0){\strut{} 0.4}}%
\put(10325,1100){\makebox(0,0){\strut{} 0.6}}%
\put(13125,1100){\makebox(0,0){\strut{} 0.8}}%
\put(15925,1100){\makebox(0,0){\strut{} 1}}%
\put(550,8650){\rotatebox{90}{\makebox(0,0){\strut{}$\bar P(n)$ or $n/N$}}}%
\put(8925,275){\makebox(0,0){\strut{}Normalized number of peers $n/N$}}%
\put(3605,13550){\makebox(0,0)[l]{\strut{}\large{\textsf{Growth Phase}}}}%
\put(10045,5850){\makebox(0,0)[l]{\strut{}\large{\textsf{Shrinkage Phase}}}}%
\put(6125,6270){\rotatebox{45}{\makebox(0,0)[l]{\strut{}\large{\textsf{Phase Boundary}}}}}%
\end{picture}%
\endgroup

}
\caption{Phase diagram of club dynamics with direction field}
\label{figure:club_direction_field}
\end{center}
\end{figure}

It is important to note that the above analysis is of the average case. The
actual dynamics of a realization of the club membership over time as
$\mathcal{C}(t)\subset \mathcal{N}$ would sketch a sample path
$(|\mathcal{C}(t)|,P_{\mathcal{C}(t)}(n))$ that staggers around the
corresponding $\bar P(n)$ curve\footnote{The staggering, or departure from
the average case, would depend on the extent and rate of mixing, viz., the
stochasticity of the club membership. Generally speaking, a large number of
active peers with strong flows both in and out of the club would stay close
to the average case with less staggering. Otherwise a sample path may
actually get stuck with a niche self-sufficient club that sees neither peers
joining nor members leaving.}. However, the family of $\bar P(n)$ curves for
all $\pi$ values define a direction field of average directions of the forces
that act upon any sample path. The average direction is towards growth above
the $n/N$ diagonal, and towards shrinkage below, as shown in figure
(\ref{figure:club_direction_field}). In other words, the $n/N$ diagonal is a
boundary between two phases of the club dynamics, a growth phase for the club
states above it and a shrinkage phase for those below. This is a powerful way
to visualize the club dynamics, especially when $\pi$ may vary over time in
more complex cases.
%20050226

\subsection{Critical behaviour and multiple equilibria}

%20050226
Note that $p_i(0)=0$ and $p_i(n)$ is bounded and concave increasing in $n$.
When $d=1$, $\bar P(n)$ is bounded and concave increasing in $n$ also.
Subsequently, there is at most one stable positive fixed point.  Theorem
\ref{theorem:instability} establishes a sharp threshold for $\pi$, a
composite \emph{control parameter} of the club as a dynamical system. The
club would stabilize at an empty membership when $\pi < 1$, or the unique
stable positive fixed point of equation (\ref{eqn:dynamic_equilibrium})
otherwise. In cases when $\pi$ varies across the threshold of unity, the club
would undergo critical change,
 and move towards either of the two stable fixed points.\\

\begin{figure}[htbp]
\begin{center}
{\scriptsize
\begin{picture}(0,0)%
\includegraphics{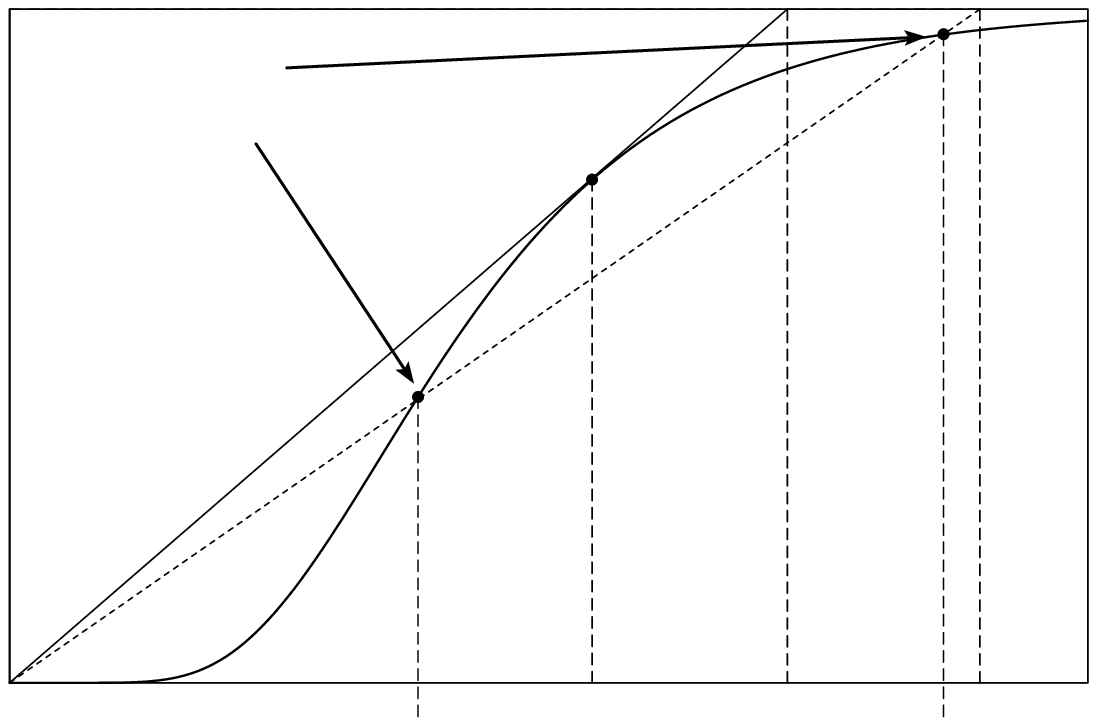}%
\end{picture}%
\begingroup
\setlength{\unitlength}{0.0200bp}%
\begin{picture}(18000,10800)(0,0)%
\put(550,5400){\rotatebox{90}{\makebox(0,0){\strut{}$\bar P(n)$ or $n/N$}}}%
\put(2759,9377){\makebox(0,0)[l]{\strut{}two fixed points}}%
\put(2759,8504){\makebox(0,0)[l]{\strut{}beside $n_{crit}$}}%
\put(7477,-321){\makebox(0,0)[l]{\strut{}$\frac{\partial \bar P}{\partial n} > \frac{1}{N}$}}%
\put(15100,-321){\makebox(0,0)[l]{\strut{}$\frac{\partial \bar P}{\partial n} < \frac{1}{N}$}}%
\put(10043,259){\makebox(0,0)[l]{\strut{}$n_{crit}$}}%
\put(5642,1520){\makebox(0,0)[l]{\strut{}$\bar P(n)$}}%
\put(12739,259){\makebox(0,0)[l]{\strut{}$N_{crit}$}}%
\put(15623,259){\makebox(0,0)[l]{\strut{}$N$}}%
\put(985,10250){\makebox(0,0)[l]{\strut{}$1$}}%
\put(985,356){\makebox(0,0)[l]{\strut{}$0$}}%
\put(7477,-903){\makebox(0,0)[l]{\strut{}unstable}}%
\put(15100,-903){\makebox(0,0)[l]{\strut{}stable}}%
\end{picture}%
\endgroup
    \vspace{5mm}
}\caption{Critical population and bifurcation of fixed points, for $d>1$}
\label{figure:s_shape}
\end{center}
\end{figure}

When $d>1$, an empty membership is always stable according to Theorem \ref{theorem:instability}.
For peer population above some minimum level $N_{crit}>0$ such that $n/N_{crit}$ is first
tangential to $\bar P(n)$ as in figure (\ref{figure:s_shape}),
at least two positive fixed points exist.
The smaller one would be unstable while the larger is
always stable (see figure (\ref{figure:s_shape})).
The smaller fixed point signifies a lower threshold, a ``critical mass" of membership needed for autonomous growth thereafter.
The club would be in danger of collapse whenever its membership falls below
this level, even when such fall is transient to begin with.

%20050226

\begin{proposition}[Critical Population and Bifurcation]\label{prop:d_greater_than_1}
$N_{crit}$ is the smallest solution to the simultaneous equations
$$
\left.\frac{\partial \bar P(n)}{\partial n}\right|_{n_{crit}}
= \frac{\bar P(n_{crit})}{n_{crit}} = \frac{1}{N_{crit}}
$$
where $n_{crit}$ is a bifurcation point: once $N$ increases above $N_{crit}$,
two fixed points appear on either sides of $n_{crit}$ and move away from it.
\end{proposition}

This proposition follows simply from the fact that $\bar P(n)$ is smooth,
increasing and upper bounded at $1$ (see figure (\ref{figure:s_shape})). The
membership level $n_{crit}$ is metastable as it is exactly marginal to the
stability condition (\ref{eqn:start_up_condition}). In the special case when
the peers are not differentiated in that $h_i(s) = h(s)$, the increase in
$\bar P(n)$ concentrates around an inflection point just before $n_{crit}$.
However, when the $h_i(s)$'s are spread out so that $\sum_s h(s) g(s)$ is
highly variable, $\bar P(n)$ would increase more gradually. As a result, the
bifurcation may occur more sharply with a wider spread between the two
resulting fixed points.

%%%%%%
\section{A numerical example with truncated Zipfian aggregate \mbox{demand}}

Consider a population of $N$ peers with a truncated Zipfian
aggregate supply, viz.:
\begin{equation}
g(s) = c s^{-\beta} \;\;\;\; 1\le s\le s_{max} \label{eqn:example_zipf}
\end{equation}
where $c=(\sum_{s=1}^{s_{max}}  s^{-\beta})^{-1}$.
This rank-frequency distribution is widely observed in Web and
\mbox{peer-to-peer} file popularity measurement studies \cite{gdsglz2003}, \cite{sgg2002} .
The exponent $\beta$ is often around and below $1$.
Its skewness as measured by its norm is
$$
\Vert g \Vert=  c\,\sqrt{\sum_{s=1}^{s_{max}}s^{-2\beta}}
$$
which is determined by two key parameters, viz. the \emph{peakedness}
of the Zipfian distribution as governed by the exponent $\beta$,
and the \emph{variety} of chunk types as governed by $s_{max}$.\\

Generally speaking, $g(s)$ may match the aggregate demand $h(s)$ to different
degrees.  Below we analyze two cases, viz. the perfect match case when
$h(s)=g(s)$ and the imperfect match case due to a simple shift between $h(s)$
and $g(s)$.  Also, we consider simple requests ($d=1$) throughout.

\subsection{Perfect match case: $h(s)=g(s)$}
\begin{figure}[h]
\begin{center}
{\scriptsize
\begin{picture}(0,0)%
\includegraphics{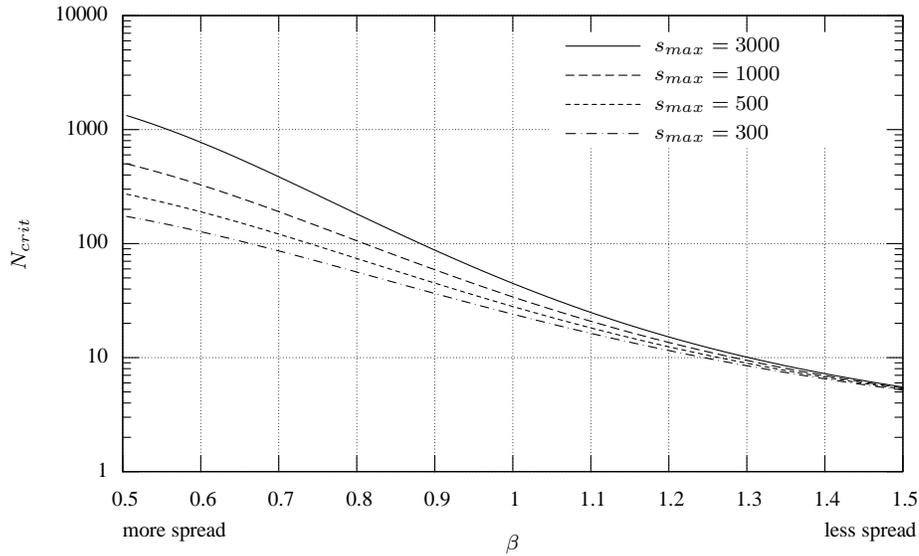}%
\end{picture}%
\begingroup
\setlength{\unitlength}{0.0200bp}%
\begin{picture}(18000,10800)(0,0)%
\put(2200,1650){\makebox(0,0)[r]{\strut{} 1}}%
\put(2200,3800){\makebox(0,0)[r]{\strut{} 10}}%
\put(2200,5950){\makebox(0,0)[r]{\strut{} 100}}%
\put(2200,8100){\makebox(0,0)[r]{\strut{} 1000}}%
\put(2200,10250){\makebox(0,0)[r]{\strut{} 10000}}%
\put(2475,1100){\makebox(0,0){\strut{} 0.5}}%
\put(3945,1100){\makebox(0,0){\strut{} 0.6}}%
\put(5415,1100){\makebox(0,0){\strut{} 0.7}}%
\put(6885,1100){\makebox(0,0){\strut{} 0.8}}%
\put(8355,1100){\makebox(0,0){\strut{} 0.9}}%
\put(9825,1100){\makebox(0,0){\strut{} 1}}%
\put(11295,1100){\makebox(0,0){\strut{} 1.1}}%
\put(12765,1100){\makebox(0,0){\strut{} 1.2}}%
\put(14235,1100){\makebox(0,0){\strut{} 1.3}}%
\put(15705,1100){\makebox(0,0){\strut{} 1.4}}%
\put(17175,1100){\makebox(0,0){\strut{} 1.5}}%
\put(550,5950){\rotatebox{90}{\makebox(0,0){\strut{}$N_{crit}$}}}%
\put(9825,275){\makebox(0,0){\strut{}$\beta$}}%
\put(2475,526){\makebox(0,0)[l]{\strut{}more spread}}%
\put(15705,526){\makebox(0,0)[l]{\strut{}less spread}}%
\put(12500,9675){\makebox(0,0)[l]{\strut{}$s_{max} = 3000$}}%
\put(12500,9125){\makebox(0,0)[l]{\strut{}$s_{max} = 1000$}}%
\put(12500,8575){\makebox(0,0)[l]{\strut{}$s_{max} = 500$}}%
\put(12500,8025){\makebox(0,0)[l]{\strut{}$s_{max} = 300$}}%
\end{picture}%
\endgroup
}\caption{$N_{crit}$ vs ($\beta$, $s_{max}$) for perfect match case ($k\,\rho=1$)}
\label{figure:Ncrit_vs_zipf_beta}
\end{center}
\end{figure}

\begin{figure}[!h]
\begin{center}
{\scriptsize
    \begin{picture}(0,0)%
\includegraphics{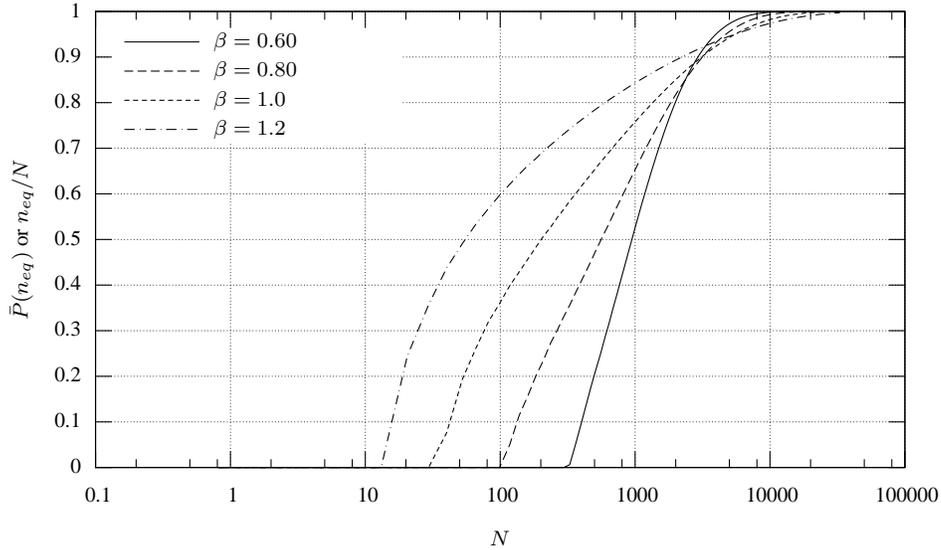}%
\end{picture}%
\begingroup
\setlength{\unitlength}{0.0200bp}%
\begin{picture}(18000,10800)(0,0)%
\put(1650,1650){\makebox(0,0)[r]{\strut{} 0}}%
\put(1650,2510){\makebox(0,0)[r]{\strut{} 0.1}}%
\put(1650,3370){\makebox(0,0)[r]{\strut{} 0.2}}%
\put(1650,4230){\makebox(0,0)[r]{\strut{} 0.3}}%
\put(1650,5090){\makebox(0,0)[r]{\strut{} 0.4}}%
\put(1650,5950){\makebox(0,0)[r]{\strut{} 0.5}}%
\put(1650,6810){\makebox(0,0)[r]{\strut{} 0.6}}%
\put(1650,7670){\makebox(0,0)[r]{\strut{} 0.7}}%
\put(1650,8530){\makebox(0,0)[r]{\strut{} 0.8}}%
\put(1650,9390){\makebox(0,0)[r]{\strut{} 0.9}}%
\put(1650,10250){\makebox(0,0)[r]{\strut{} 1}}%
\put(1925,1100){\makebox(0,0){\strut{} 0.1}}%
\put(4467,1100){\makebox(0,0){\strut{} 1}}%
\put(7008,1100){\makebox(0,0){\strut{} 10}}%
\put(9550,1100){\makebox(0,0){\strut{} 100}}%
\put(12092,1100){\makebox(0,0){\strut{} 1000}}%
\put(14633,1100){\makebox(0,0){\strut{} 10000}}%
\put(17175,1100){\makebox(0,0){\strut{} 100000}}%
\put(550,5950){\rotatebox{90}{\makebox(0,0){\strut{}$\bar P(n_{eq})$ or $n_{eq}/N$}}}%
\put(9550,275){\makebox(0,0){\strut{}$N$}}%
\put(4150,9675){\makebox(0,0)[l]{\strut{}$\beta =0.60$}}%
\put(4150,9125){\makebox(0,0)[l]{\strut{}$\beta =0.80$}}%
\put(4150,8575){\makebox(0,0)[l]{\strut{}$\beta =1.0$}}%
\put(4150,8025){\makebox(0,0)[l]{\strut{}$\beta =1.2$}}%
\end{picture}%
\endgroup

}\caption{Expected equilibrium membership and success rate vs $N\,k\,\rho$ ($s_{max} = 1000$)}
\label{figure:zipf_P_vs_Ncrit}
\end{center}
\end{figure}
According to Theorem \ref{theorem:instability}, the autonomous growth condition is
\begin{eqnarray}
N_{crit} &\ge&  \frac{1}{k_{crit} \;\rho_{crit}} \;\frac{1}{c^2\sum_s s^{-2\beta}} \label{eqn:numeric_zipf_1}
\end{eqnarray}
since $\langle g,h \rangle=1$ and $\Vert g\Vert = \Vert h
\Vert=c\,\sqrt{\sum_{s}s^{-2\beta}}$. The dependence of $N_{crit}$ on $\beta$
and $s_{max}$ is shown in figure (\ref{figure:Ncrit_vs_zipf_beta}).
Autonomous growth is favoured by large $\beta$ (peakedness) and low $s_{max}$
(variety).

Once the control parameter of the club is above the growth threshold,
it would sustain around an equilibrium membership size $n_{eq}$ as the unique
stable fixed point of equation (\ref{eqn:dynamic_equilibrium}). Solving for
different values of $Nk\rho$ and $\beta$ gives figure
(\ref{figure:zipf_P_vs_Ncrit}). This figure shows the proportion $n_{eq}/N$,
which is also the performance level of the club in terms of $\bar P(n_{eq})$,
the average success rate of information search in the club.

\subsection{Imperfect match due to simple shift}\label{sec:match}
Supply and demand distributions would seldom match perfectly.  In fact, one
would often be considered leading the other.  For example, a \emph{demand
lead} case would demand a wider variety of goods than the aggregate supply
distribution offers, while a \emph{supply lead} case sees more variety in the
supplied goods. The ``excess'' in demanded types (or supplied types) reflects
the types of goods that the supply (or the demand) cannot follow at a
particular moment.  Here we consider all such excess being concentrated in
either the lowest or the highest ranks for simple illustrations.\\

\begin{figure}[hbtp]
\begin{center}
{\scriptsize
\hbox{
\begin{picture}(0,0)%
\includegraphics{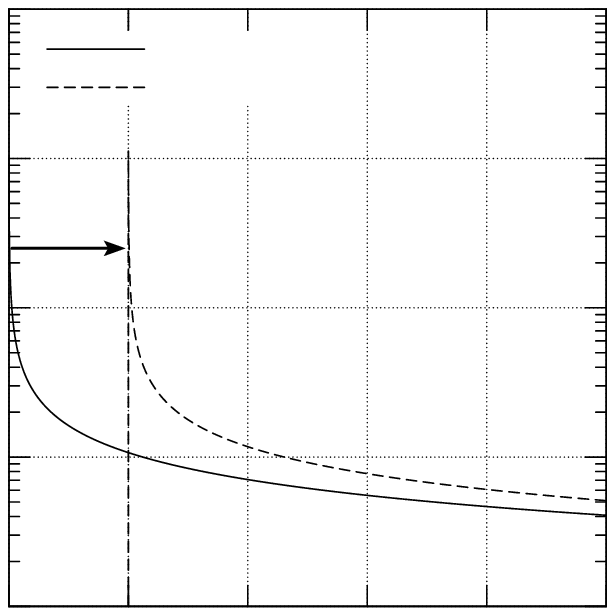}%
\end{picture}%
\begingroup
\setlength{\unitlength}{0.0200bp}%
\begin{picture}(10800,10800)(0,0)%
\put(2475,1650){\makebox(0,0)[r]{\strut{} 1e-005}}%
\put(2475,3800){\makebox(0,0)[r]{\strut{} 0.0001}}%
\put(2475,5950){\makebox(0,0)[r]{\strut{} 0.001}}%
\put(2475,8100){\makebox(0,0)[r]{\strut{} 0.01}}%
\put(2475,10250){\makebox(0,0)[r]{\strut{} 0.1}}%
\put(2749,1100){\makebox(0,0){\strut{} 0}}%
\put(4469,1100){\makebox(0,0){\strut{} 2000}}%
\put(6189,1100){\makebox(0,0){\strut{} 4000}}%
\put(7910,1100){\makebox(0,0){\strut{} 6000}}%
\put(9630,1100){\makebox(0,0){\strut{} 8000}}%
\put(11350,1100){\makebox(0,0){\strut{} 10000}}%
\put(7050,275){\makebox(0,0){\strut{}$s$ rank}}%
\put(3007,6976){\makebox(0,0)[l]{\strut{}\tiny{$\delta = 2000$}}}%
\put(4975,9675){\makebox(0,0)[l]{\strut{}$g(s)$}}%
\put(4975,9125){\makebox(0,0)[l]{\strut{}$h(s)$}}%
\end{picture}%
\endgroup
%%%%%%%%%%%%%%%%%%%%%%%%%%%%%%%%%%%%%%%%%%%%%%%%%%%%%%%%%%%%%

        \begin{picture}(0,0)%
\includegraphics{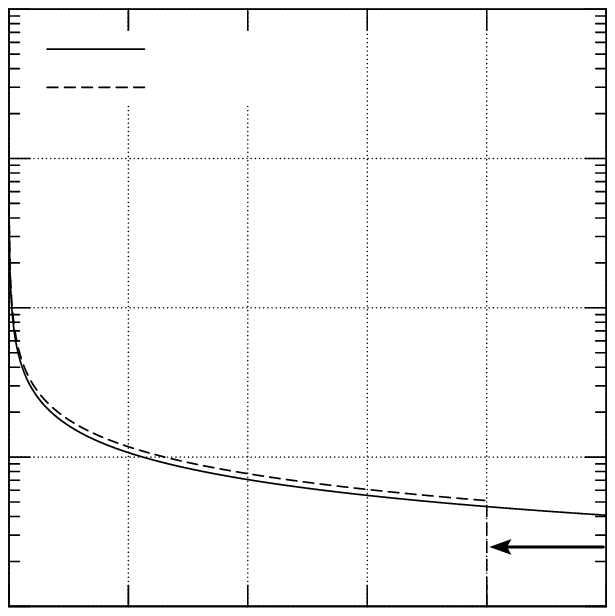}%
\end{picture}%
\begingroup
\setlength{\unitlength}{0.0200bp}%
\begin{picture}(10800,10800)(0,0)%
\put(2475,1650){\makebox(0,0)[r]{\strut{} 1e-005}}%
\put(2475,3800){\makebox(0,0)[r]{\strut{} 0.0001}}%
\put(2475,5950){\makebox(0,0)[r]{\strut{} 0.001}}%
\put(2475,8100){\makebox(0,0)[r]{\strut{} 0.01}}%
\put(2475,10250){\makebox(0,0)[r]{\strut{} 0.1}}%
\put(2749,1100){\makebox(0,0){\strut{} 0}}%
\put(4469,1100){\makebox(0,0){\strut{} 2000}}%
\put(6189,1100){\makebox(0,0){\strut{} 4000}}%
\put(7910,1100){\makebox(0,0){\strut{} 6000}}%
\put(9630,1100){\makebox(0,0){\strut{} 8000}}%
\put(11350,1100){\makebox(0,0){\strut{} 10000}}%
\put(7050,275){\makebox(0,0){\strut{}$s$ rank}}%
\put(9673,2820){\makebox(0,0)[l]{\strut{}\tiny{$\delta = -2000$}}}%
\put(4975,9675){\makebox(0,0)[l]{\strut{}$g(s)$}}%
\put(4975,9125){\makebox(0,0)[l]{\strut{}$h(s)$}}%
\end{picture}%
\endgroup
        }
       \hbox{\hspace{48mm}\hbox{(a)\hspace{74mm}(b)}}
        }

\caption{A supply lead case example ($s_{smax} = 10000$) with (a) positive $\delta$ and (b) negative $\delta$}
\label{figure:delta_meaning}
\end{center}
\end{figure}

In the supply lead case, the supply $g(s)$ is the same as defined in equation
(\ref{eqn:example_zipf}), and the demand distribution is :
\begin{displaymath}
h(s) = \left\{ \begin{array}{ll}
    0 & \textrm{if $s \le \delta$}\\
      c'(s-\delta)^{-\beta} &\textrm{if $\delta < s \le s_{max}$}
    \end {array}
\right.
\end{displaymath}
for $\delta \ge 0$, and
\begin{displaymath}
h(s) = \left\{ \begin{array}{ll}
    c{''}s^{-\beta} & \textrm{if $s \le s_{max} + \delta$}\\
     0 &\textrm{if $ s_{max} + \delta < s \le s_{max}$}
    \end {array}
\right.
\end{displaymath}
for $\delta <0$. $c'$ and $c{''}$ are normalizing constants such that $\sum_s
h(s) = 1$. See figure (\ref{figure:delta_meaning}) for an illustrated
example. A positive shift $\delta>0$ means the excess types occupy the
highest ranks, while a negative shift means they occupy the lowest ranks. The
supply lead case would simply have the expressions of $g(s)$ and $h(s)$
exchanged.\\

\begin{figure}[hbtp]
\begin{center}
{\scriptsize
\begin{picture}(0,0)%
\includegraphics{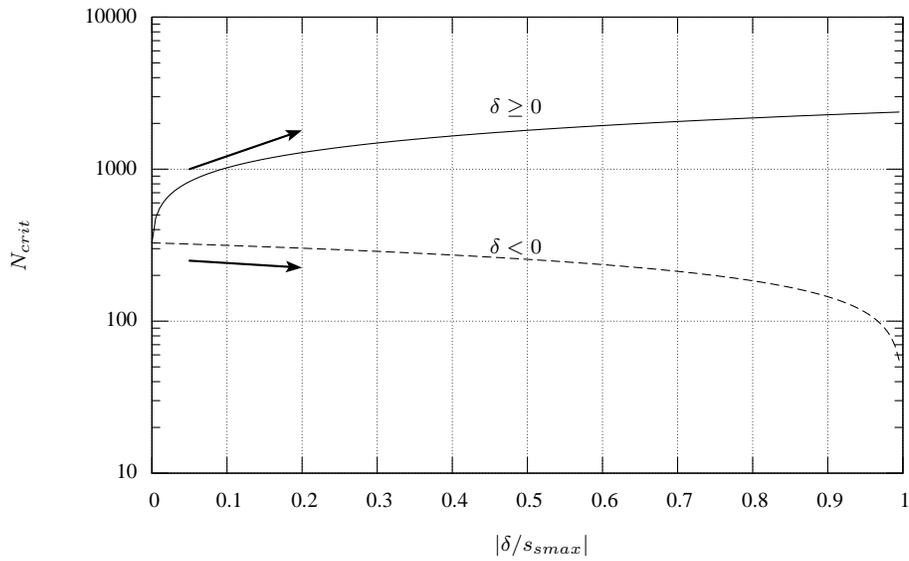}%
\end{picture}%
\begingroup
\setlength{\unitlength}{0.0200bp}%
\begin{picture}(18000,10800)(0,0)%
\put(2750,1650){\makebox(0,0)[r]{\strut{} 10}}%
\put(2750,4517){\makebox(0,0)[r]{\strut{} 100}}%
\put(2750,7383){\makebox(0,0)[r]{\strut{} 1000}}%
\put(2750,10250){\makebox(0,0)[r]{\strut{} 10000}}%
\put(3025,1100){\makebox(0,0){\strut{} 0}}%
\put(4440,1100){\makebox(0,0){\strut{} 0.1}}%
\put(5855,1100){\makebox(0,0){\strut{} 0.2}}%
\put(7270,1100){\makebox(0,0){\strut{} 0.3}}%
\put(8685,1100){\makebox(0,0){\strut{} 0.4}}%
\put(10100,1100){\makebox(0,0){\strut{} 0.5}}%
\put(11515,1100){\makebox(0,0){\strut{} 0.6}}%
\put(12930,1100){\makebox(0,0){\strut{} 0.7}}%
\put(14345,1100){\makebox(0,0){\strut{} 0.8}}%
\put(15760,1100){\makebox(0,0){\strut{} 0.9}}%
\put(17175,1100){\makebox(0,0){\strut{} 1}}%
\put(550,5950){\rotatebox{90}{\makebox(0,0){\strut{}$N_{crit}$}}}%
\put(10375,275){\makebox(0,0){\strut{}$|\delta/s_{smax}|$}}%
\put(9393,8524){\makebox(0,0)[l]{\strut{}$\delta \ge 0$}}%
\put(9393,5884){\makebox(0,0)[l]{\strut{}$\delta < 0$}}%
\end{picture}%
\endgroup
}\caption{Effect of mismatch due to simple shift ($s_{max}=1000$, $k\rho =1$, $\beta = 0.6$)}
\label{figure:Ncrit_vs_delta}
\end{center}
\end{figure}

Figure (\ref{figure:Ncrit_vs_delta}) shows that excess in the highest ranks are very demanding and would
require a very large increase in $N_{crit}$ for autonomous growth.
However, excess in the lowest ranks actually decreases $N_{crit}$ and autonomous growth becomes easier.
This suggests that focussing of supply on chunk types of the highest ranks would trigger autonomous growth more readily.\\

In summary, the distinction between supply lead and demand lead cases is immaterial to the autonomous
growth threshold, though it may be important to modelling
supply and demand dynamics. What matters is where they differ --- in the higher or lower ranks.

\section{Discussion}

The simple ISC model displays interesting behaviour resulting from coupling
of the membership and content dynamics.  As a dynamical system, it exhibits
phase transition in a composite control parameter $\pi$.  Information sharing
is sustainable with a good membership only when $\pi$ is above a threshold of
$1$. It implies no effort in increasing $\pi$ is worthwhile (e.g. by
improving efficiency and supply, or increasing peer population) unless
$\pi$ goes above the threshold as a result.\\

Many researchers study the problem of excessive free-riding \cite{fpcs2004},
\cite{hp2001}, \cite{ah2000} that corresponds to the existence of many peers
with $K_i=0$ in our model. Incentive mechanisms are devised which reward
contribution and/or penalize free-riding.  One way to analyze such mechanisms
is by extending the ISC model so that the search efficiency a peer sees
becomes an increasing function of the contribution he decides on a rational
basis. In this case, our ongoing study indicates that empty membership may
become stable always, even for $d=1$, as long as the club relies solely on
members for content.  An empty club and a rational (frugal) peer population
are in a dead lock situation.  The dead lock may be broken only when either
the club has sufficient initial content to attract the peers, or some peers
are generous enough to contribute without expecting extra
benefit (therefore not rational in the conventional sense).\\

However, free-riding is not a problem \emph{per se} under the non-rivalrous
assumption.  As free-riders are those with $K_i=0$, we may redefine $N$ to
exclude them so that only contributing peers (those with positive payloads)
are counted.  As a result, $N$ is reduced, and the average payload size
$k$ is increased (while the club's average content $Nk$ remains the same as
before).  All results
in this paper would continue to hold, albeit with the notion of a peer redefined.
What matters is the contributing peer population: the club would grow as
long as they are joining to share enough content and attract sufficiently
many of themselves.  The existence of free-riders is phantom to the system.
They would be no nuisance as long as provisioning of extra copies carries
no sharing cost.\\

Incentives would help not by reducing free-riding but by increasing the
contributing peer population, viz. $N$.  The distinction may seem frivolous
as reduced free-riding is often regarded as increased contribution.  However
this may not be true always.  One may imagine a negative incentive scheme
which merely causes free-riders to demand less, or turn away altogether,
without turning them into contributing peers.  The club's content is
not benefited.  As incentive schemes are often costly to maintain in
practice, negative schemes as such should be saved for positive ones
that aim to increase $N$ directly.  A reasonable principle in economizing the
use of incentive schemes would be to focus on those peers who are
bordering on free-riding, to coerce them into contributing.\\

In fact, the club's well being may actually be harmed in two possible ways
when free-riding is discouraged.  First, free-riders may develop into
contributors if only they stay long enough for the club to become
sufficiently important to them.  Second, they may in fact be useful audience
to the members, e.g. in newsgroups, BBS and forums, where wider circulation
of the shared information often improves \emph{all} due to network effects.
(This would be diametrically \emph{opposite} to the rivalrous assumption, and
could actually be more appropriate than the non-rivalrous assumption if it
more than offsets any sharing costs due to rivalrous consumption of other
club resources.)

In cases where the non-rivalrous assumption is not appropriate due to
significant sharing costs, e.g. in processing, storage and/or network
bandwidth, penalizing free-riding would be more necessary to reduce
their loading on the system and the contributing peers.
A possible corresponding extension of the ISC model is to incorporate the
natural reduction in availability of information goods as their demand
increases.  For instance, the failure rate of demand for chunk type $s$ may
become an increasing function of its total demand $nh(s)$ one way or the
other. However, the choice of functional relation between availability and
demand should depend on the nature of the sharing cost.

Apart from extensions needed in rivalrous situations,
the ISC model has two intrinsic limitations.  First, it captures
only the average case behaviour of a nonlinear and stochastic dynamical
system.  Transient and lock-in, especially when the club is small and
peers act with large delay, may render the average case view totally useless.
Second, the join/leave
decisions are often more heterogeneous than assumed here. Requests may comprise
variable numbers of demand instances and peers may deliberate their decisions and behave differently.

\section{Conclusion and further works}
We have analyzed information sharing in a very general setting, by means of a
statistical model (ISC) with peers of different demand and supply of
information.  As a dynamical system, the model exhibits interesting critical
behaviour with multiple equilibria.
A unique feature of the ISC model is that information is chunked and
typed, as we believe modelling the composition of the information
content is crucial in many situations of interest.  Subsequently, it
displays a sharp growth threshold that depends on the goodness of match
in the types of information being demanded and supplied by the sharing
members.  While being rich in behaviour, this model is simple enough for
detail analysis of the equilibrium states.  In particular, we analyzed a
truncated Zipfian distribution of information types and derived the
growth threshold for the existence of any sustainable equilibrium, as well
as the corresponding membership size and performance level.\\

Much simplicity of the ISC model stems from the non-rivalrous assumption
made.  Real situations are more complicated in that peers may be sharing both
rivalrous resources and non-rivalrous information at the same time.  However,
Benkler \cite{b_sharing_2004} points out that overcapacity is a growing trend
in distributed systems such as the Internet, so much so that even rivalrous
resources are increasingly being shared like non-rivalrous goods.  On the
other hand, free-riding would work in the opposite direction if the community
in question is prosperous and attracts so much free-riding that contention
for some shared rivalrous resources begins to happen.  The challenge is then
to identify the major sources of social cost of sharing \cite{v2003} and
properly account for them. A natural extension of our work would be to study
the interplay between an information sharing community and different types of
host networks.  Another extension is certainly the incentive issue: how
incentive schemes should be devised in response to different mixes of
rivalrous and non-rivalrous resources.
%The ongoing study we mentioned in passing begins with a
%distribution of peer types according to a measure of their propensity to
%free-ride, in terms of the marginal rate of their private cost-benefit change in contributing to the community.

\bibliographystyle{ieeetran}
\bibliography{all}

\begin{thebibliography}{10}
\providecommand{\url}[1]{#1}
\csname url@rmstyle\endcsname
\providecommand{\newblock}{\relax}
\providecommand{\bibinfo}[2]{#2}
\providecommand\BIBentrySTDinterwordspacing{\spaceskip=0pt\relax}
\providecommand\BIBentryALTinterwordstretchfactor{4}
\providecommand\BIBentryALTinterwordspacing{\spaceskip=\fontdimen2\font plus
\BIBentryALTinterwordstretchfactor\fontdimen3\font minus
  \fontdimen4\font\relax}
\providecommand\BIBforeignlanguage[2]{{%
\expandafter\ifx\csname l@#1\endcsname\relax
\typeout{** WARNING: IEEEtran.bst: No hyphenation pattern has been}%
\typeout{** loaded for the language `#1'. Using the pattern for}%
\typeout{** the default language instead.}%
\else
\language=\csname l@#1\endcsname
\fi
#2}}

\bibitem{SETI}
\BIBentryALTinterwordspacing
``{SETI@home}.'' [Online]. Available: \url{http://setiathome.ssl.berkeley.edu/}
\BIBentrySTDinterwordspacing

\bibitem{ll2002}
J.~Y.~B. Lee and W.~T. Leung, ``Design and analysis of a fault-tolerant
  mechanism for a server-less video-on-demand system,'' in \emph{Proceedings of
  International Conference on Parallel and Distributed Systems}, 2002.

\bibitem{hhbb2003}
M.~Hefeeda, A.~Habib, B.~Botev, D.~Xu, and B.~Bhargava, ``{PROMISE}:
  Peer-to-peer media streaming using collectcast,'' in \emph{Proceedings of ACM
  Multimedia}, 2003.

\bibitem{kazza}
\BIBentryALTinterwordspacing
``Kazaa.'' [Online]. Available: \url{http://www.kazaa.com/}
\BIBentrySTDinterwordspacing

\bibitem{bittorrent}
\BIBentryALTinterwordspacing
``Bittorrent.'' [Online]. Available: \url{http://bitconjurer.org/BitTorrent/}
\BIBentrySTDinterwordspacing

\bibitem{gj2003}
B.~Gu and S.~Jarvenpaa, ``Are contributions to {P2P} technical forums private
  or public goods? - an empirical investigation,'' in \emph{Proceedings of
  Workshop on Economics of {P2P} Systems}, June 2003.

\bibitem{kstt2002}
R.~Krishnan, M.~D. Smith, Z.~Tang, and R.~Telang, ``The virtual commons: Why
  free-riding can be tolerated in file sharing networks,'' in \emph{Proceedings
  of {International Conference on Information Systems}}, 2002.

\bibitem{bas2003}
C.~Buragohain, D.~Agrawal, and S.~Suri, ``A game theoretic framework for
  incentives in {P2P} systems,'' in \emph{Proceedings of the Third
  International Conference on Peer-to-Peer Computing}, 2003.

\bibitem{gbml2001}
P.~Golle, K.~Leyton-Brown, I.~Mironov, and M.~Lillibridge, ``Incentives for
  sharing in peer-to-peer networks,'' in \emph{Proceedings of the 2001 ACM
  Conference on Electronic Commerce}, 2001.

\bibitem{rrsf2003}
K.~Ranganathan, M.~Ripeanu, A.~Sarin, and I.~Foster, ``To share or not to
  share: An analysis of incentives to contribute in collaborative file sharing
  environments,'' in \emph{Proceedings of Workshop on Economics of {P2P}
  Systems}, June 2003.

\bibitem{fpcs2004}
M.~Feldman, C.~Papadimitriou, J.~Chuang, and I.~Stoica, ``Free-riding and
  whitewashing in peer-to-peer systems,'' in \emph{Proceedings of ACM SIGCOMM
  Workshop on Practice and Theory of Incentives in Networked Systems}, 2004.

\bibitem{gdsglz2003}
K.~P. Gummadi, R.~J. Dunn, S.~Saroiu, S.~D. Gribble, H.~M. Levy, and
  J.~Zahorjan, ``Measurement, modeling, and analysis of a peer-to-peer
  file-sharing workload,'' in \emph{Proceedings of Symposium on Operating
  Systems Principles}, 2003.

\bibitem{sgg2002}
S.~Saroiu, P.~K. Gummadi, and S.~D. Gribble, ``A measurement study of
  peer-to-peer file sharing systems,'' in \emph{Proceedings of Multimedia
  Computing and Networking}, 2002.

\bibitem{hp2001}
J.~Hindriks and R.~Pancs, ``Free riding on altruism and group size,'' Queen
  Mary College, University of London, Department of Economics, Tech. Rep.
  wp-436, 2001.

\bibitem{ah2000}
E.~Adar and B.~Huberman, ``Free riding on gnutella,'' \emph{First Monday},
  vol.~5, Sep 2000.

\bibitem{b_sharing_2004}
Y.~Benkler, ``{Sharing nicely}: on shareable goods and the emergence of sharing
  as a modality of economic production,'' \emph{The Yale Law Journal}, vol.
  114, pp. 273--358, 2004.

\bibitem{v2003}
H.~R. Varian, ``The social cost of sharing,'' in \emph{Proceedings of Workshop
  on Economics of {P2P} Systems}, June 2003.

\end{thebibliography}

\begin{biography}{\includegraphics[width=1in,height=1.25in,clip,keepaspectratio]{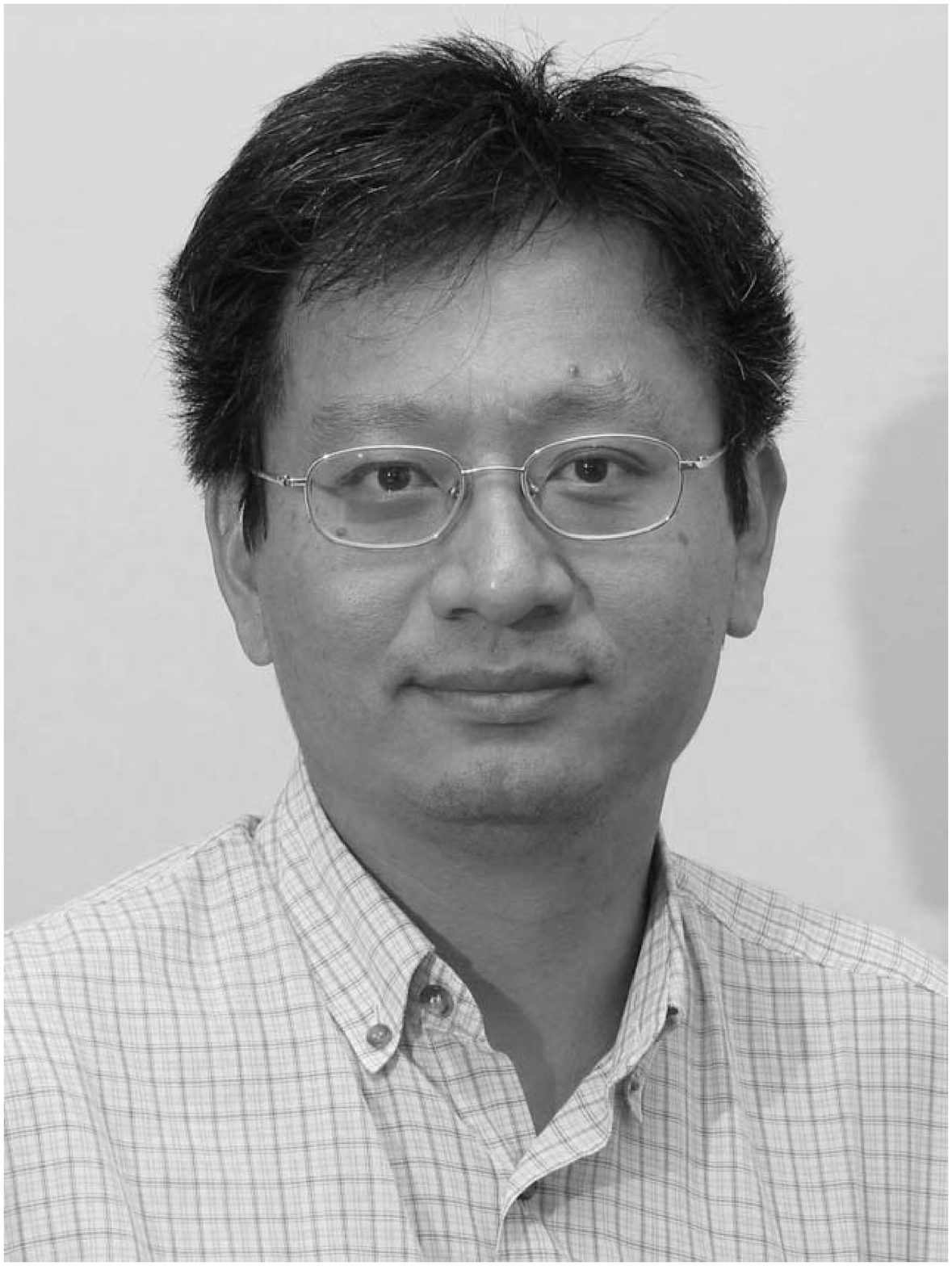}}{W.-Y. Ng}
Wai-Yin Ng received his BA in 1985 (specializing in control and
operational research) and PhD in control engineering in 1989, both from
the University of Cambridge, U.K.  and is currently associate professor
in information engineering in The Chinese University of Hong Kong.  His
current research focus is in complex networks, a young vibrant science
concerned with connectivity, complexity and emergent phenomena in both
natural and artificial systems.
\end{biography}

\begin{biography}{\includegraphics[width=1in,height=1.25in,clip,keepaspectratio]{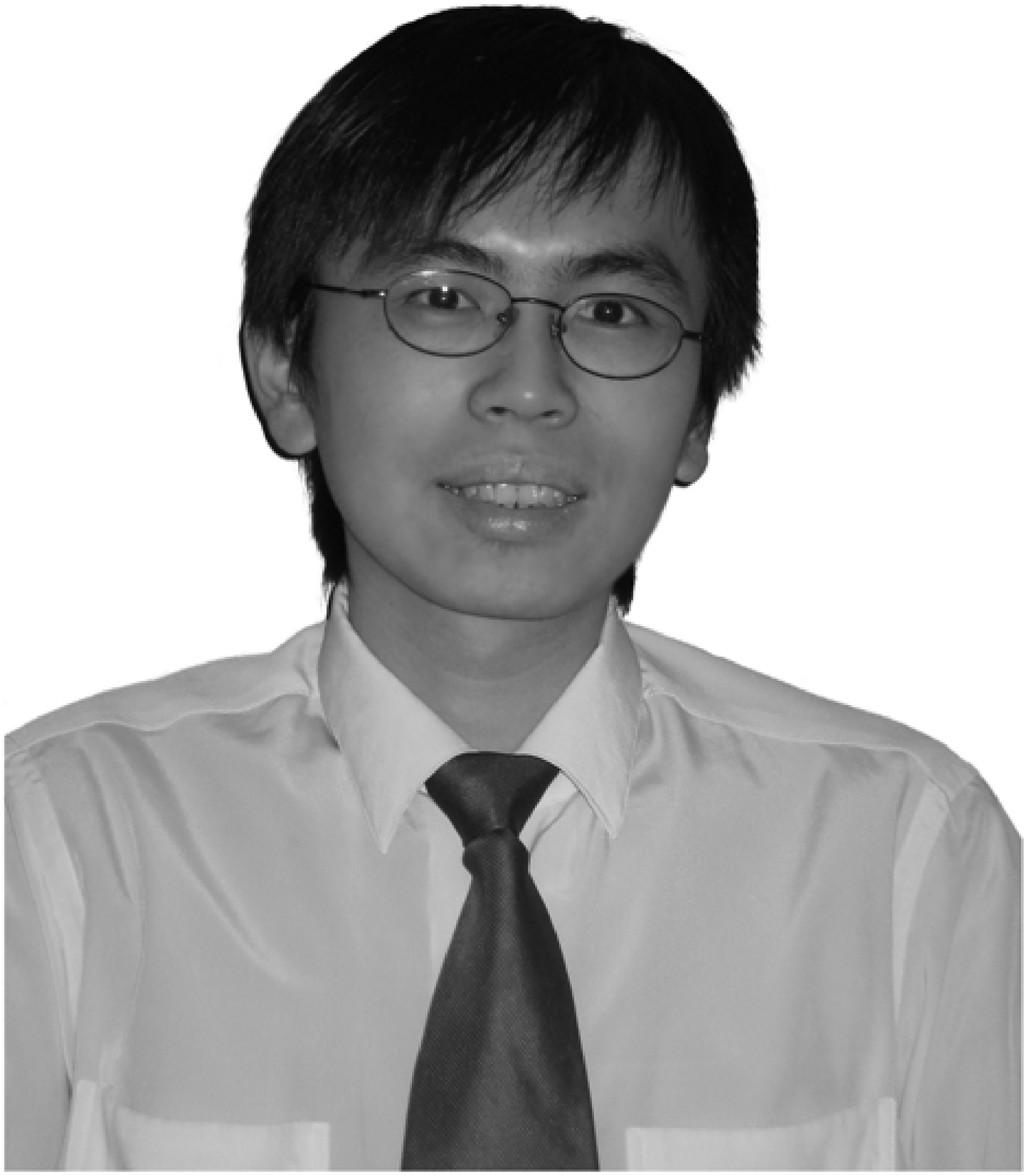}}{W.K. Lin}
Wing Kai Lin received his B.Eng degree in
2001 and is currently completing his Master degree in information engineering, both 
from the Chinese Univeristy of Hong Kong. His research interest includes 
replication in peer to peer systems and economics issues in incentive mechanisms.
\end{biography}
\vspace{0.85cm}
\begin{biography}{\includegraphics[width=1in,height=1.25in,clip,keepaspectratio]{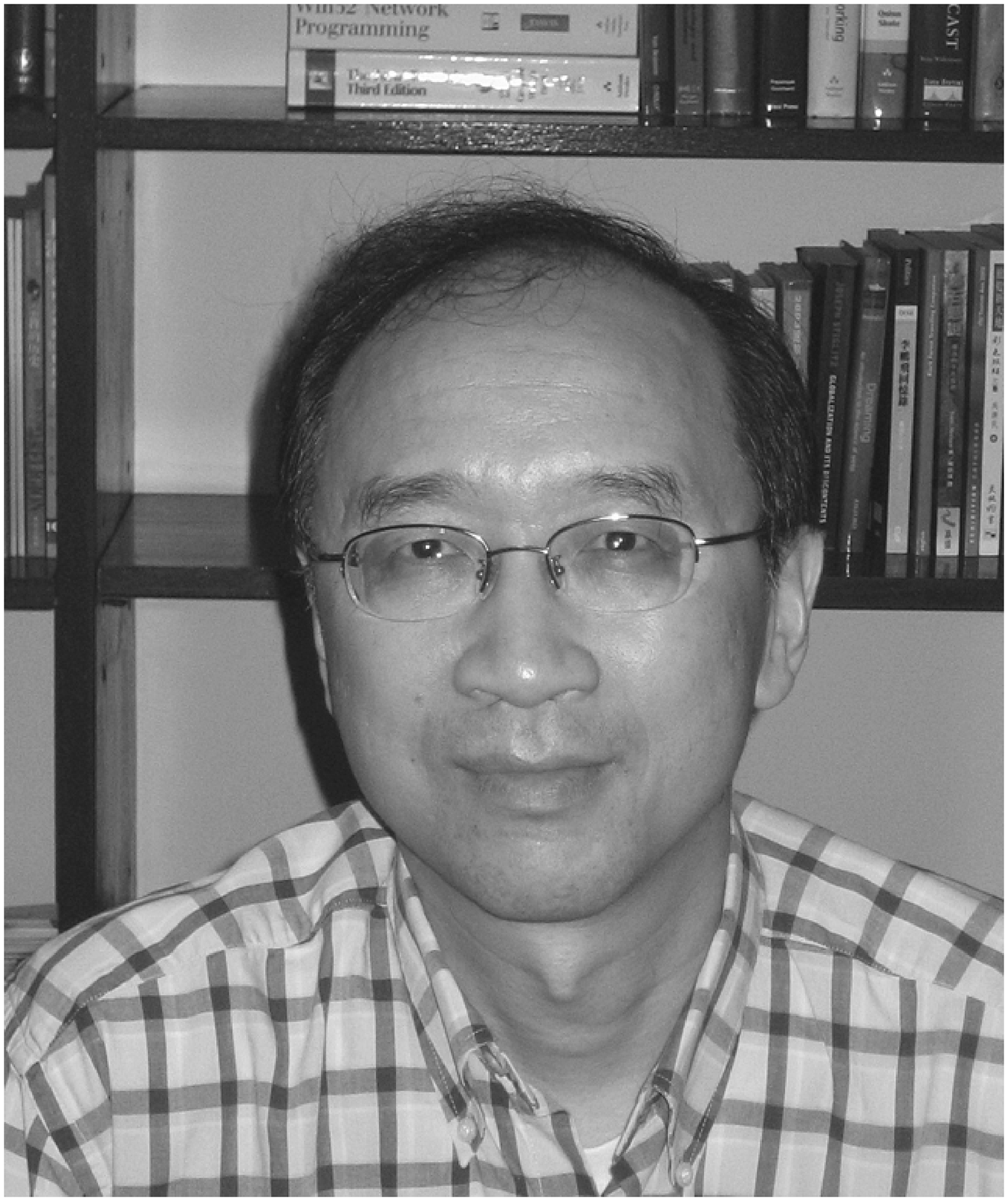}}{D.M. Chiu}
Dah Ming Chiu received his B.Sc. degree from Imperial College London in 1975,
and Ph.D. degree from Harvard University in 1980.  He worked for Bell Labs,
Digital Equipment Corporation and Sun Microsystems Laboratories. Currently,
he is a professor in the Department of Information Engineering at the
Chinese University of Hong Kong. He is on the editorial board of the
International Journal of Communication Systems.
\end{biography}

\end{document}